\documentclass[12pt]{article}
\usepackage{tabularx}
\usepackage{array}
\usepackage{graphics}
\usepackage{graphicx}
\usepackage{epsfig}
\usepackage{amsmath}
\usepackage{amssymb}
\makeatletter

\usepackage{verbatim}

\newcommand{\bea}{\begin{eqnarray}}
\newcommand{\eea}{\end{eqnarray}}
\newcommand{\beq}{\begin{equation}}
\newcommand{\eeq}{\end{equation}}
\newcommand{\gev}{{\rm GeV}}

\newcommand{\pdir}{p\kern -5.2pt\raise 0.2ex\hbox {/}}
\newcommand{\vdir}{v\kern -5.75pt\raise 0.15ex\hbox {/}}
\newcommand{\kdir}{k\kern -5.75pt\raise 0.15ex\hbox {/}}
\newcommand{\epsdir}{\epsilon\kern -5.0pt\raise 0.15ex\hbox {/}}
\newcommand{\bvdir}{\bar{v}\kern -5.75pt\raise 0.15ex\hbox {/}}
\newcommand{\Ddir}{D\kern -7.75pt\raise 0.20ex\hbox {/}}
\newcommand{\Adir}{A\kern -7.75pt\raise 0.20ex\hbox {/}}
\newcommand{\ldir}{l\kern -5.0pt\raise 0.2ex\hbox{/}}
\newcommand{\varepsdir}{\varepsilon\kern -5.5pt\raise 0.15ex\hbox{/}}

\makeatother

\begin{document}
\thispagestyle{empty} 

\begin{flushright}
\begin{tabular}{l}
{\tt LPT Orsay 02-01}\\
{\tt ROMA-1330/02}\\
{\tt UHU-FT/02-22}\\
\end{tabular}
\end{flushright}
\vskip 2.2cm\par
\begin{center}
{\par\centering \textbf{\LARGE First lattice QCD estimate} }\\
\vskip .45cm\par
{\par\centering \textbf{\LARGE of the $g_{D^\ast D\pi}$ coupling} }\\
\vskip 0.9cm\par
{\par\centering \large  
\sc A.~Abada$^a$, D.~Be\'cirevi\'c$^b$, Ph.~Boucaud$^a$, G.~Herdoiza$^a$, 
J.P.~Leroy$^a$, A.~Le~Yaouanc$^a$, O.~P\`ene$^a$, J.~Rodr\'{\i}guez--Quintero$^c$ }
{\par\centering \vskip 0.5 cm\par}
{\par\centering \textsl{ 
$^a$~Laboratoire de Physique Th\'eorique (B\^at.210), Universit\'e de
Paris XI,\\ 
Centre d'Orsay, 91405 Orsay-Cedex, France.} \\
\vskip 0.3cm\par} 
{\par\centering \textsl{
$^b$~Dipartimento di Fisica, Universit\`a di Roma ``La Sapienza",\\ 
Piazzale Aldo Moro 2, I-00185 Rome, Italy.} \\
\vskip 0.3cm\par}
{\par\centering \textsl{
$^c$ Dpto. de F\'{\i}sica Aplicada,
E.P.S. La R\'abida, Universidad de Huelva, \\21819 Palos de la fra., Spain}
\vskip 0.3cm\par }
June 24 2002                                                           
\end{center}

\vskip 0.45cm
\begin{abstract}
We present the results of the first lattice QCD study of the strong  coupling
$g_{D^\ast D\pi}$. From our simulations in the
quenched approximation,  we obtain $g_{D^\ast D\pi} = 18.8 \pm
2.3^{+1.1}_{-2.0}$ and   $\widehat g_c = 0.67 \pm 0.08^{+0.04}_{-0.06}$.
Whereas previous theoretical  studies gave different predictions, our result
favours a large value for $\widehat g_c$.  It agrees very well with  the
recent  experimental value by CLEO. $\hat g$ varies very little with the heavy
mass and we find  in the infinite mass limit $\widehat g_\infty = 0.69(18)$. 
\end{abstract}
                 \vskip 2.2 cm 
 
\setcounter{page}{1}
\setcounter{footnote}{0}
\setcounter{equation}{0}
%%%%%%%%%%%%%%%%%%%%%%%%%%%%%%%%%%%%%%%%
%%%%%%%%%%%%%%%%%%%%%%%%%%%%%%%%%%%%%%%%
%%%%%%%%%%%%%%%%%%%%%%%%%%%%%%%%%%%%%%%%

\renewcommand{\thefootnote}{\arabic{footnote}}
\vspace*{-1.5cm}
\newpage
\setcounter{footnote}{0}
%%%%%%%%%%%  Section 1
\section{Introduction}
Recent measurement of the full width of the charged $D^\ast$-meson, 
$\Gamma(D^{\ast +}) = 96\pm 4\pm 22$~keV~\cite{CLEO}, allowed for the 
experimental determination of the strong coupling of $D$-mesons to 
the $P$-wave pion $g_{D^{\ast +} D^0\pi}$, namely
\bea \label{CLEO}
g_{D^\ast D\pi} = 17.9\pm 0.3\pm 1.9\,,\quad {\rm i.e.}~~\widehat g =
0.59 \pm 0.07 \,,
\eea
where the definition of $\widehat g$ in \cite{CLEO} is quoted below.
This coupling has been extensively studied in the
literature~\footnote{~ 
A rather exhaustive list of results for this coupling can be found 
in~\cite{AD,Singer}. To the references listed there, 
one should add also ref.~\cite{newly}.},
 with a variety of approaches :
model independent approaches, the QCD sum rules,
the quark models.

{\it Model independent} approaches~\cite{cho,melbey} have produced 
windows or bounds which have been confirmed by experiment eq. (\ref{CLEO}). By
the way a rigorous bound  $\widehat g < 1$ is set by the Adler-Weisberger sum 
rule~\cite{Dominguez,AD}. Notice that $\widehat g=1$ is the 
naive non relativistic quark model result,
to be lowered by relativistic corrections.

The various {\it QCD sum rules} have been discussed 
with much care before and after the measurement in eq. (\ref{CLEO})
and have shown a surprising convergence towards
a very low value, almost a factor two below eq. (\ref{CLEO}), 
see~\cite{Colangelo}.
In particular, 
in.~\cite{Khodjamirian}, 
a value $g_{D^\ast D\pi} = 10.5 \pm 3.0$
has been quoted. No convincing explanation
has been found for this discrepancy.

{\it Quark models}, on the contrary, can accommodate
large values. Good predictions have been
produced, prior to the experimental measure~\cite{light,AD}.
But there is a large
spectrum of predictions 0.3-0.8, corresponding
to a multiplicity of models or choice
of parameters, therefore one can
wonder whether successes are truely significant,
or rather due to a happy choice. Without
entering into details, one can answer as follows:

1) Light-front quark models have indeed
a large range of predictions 0.3-0.8, mainly
because they use free quark Dirac spinors,
which yield relativistic corrections
very sensitive to the choice of the
ill-determined light quark mass.

2) Dirac type models, on the contrary, yield naturally large values of
$\widehat g \ge 0.6$, because anyway, a large effective mass is
generated for the light quark through the potential~\cite{AD}. These
authors find $\widehat g = 0.6$. Too large values obtained in other
calculations can be corrected by a quark current renormalisation factor,
but one loses predictive power.

Before claiming that the QCD based evaluations for this coupling
 are in conflict with the
experimental value, it is important to compute this coupling by employing 
the lattice QCD simulations, as recently suggested in ref.~\cite{Roudeau}.
An exploratory lattice calculation  has been performed in ref.~\cite{UKQCD}
 but in the static limit
of the heavy quark effective theory (HQET).
Here we will directly work in QCD with relativistic propagating quarks since, 
contrary to the case of $b$-physics, currently accessible lattices
allow to accommodate the charm quark mass and therefore no heavy quark 
extrapolation is needed. This  makes the lattice study of $g_{D^\ast D\pi}$
rather clean.

 In this paper, we report the first calculation of this type 
in which we used the (improved) Wilson fermions. Our final result at $\beta=6.2$
is
\bea \label{result62}
g_{D^\ast D\pi} = 18.8 \pm 2.3^{+1.1}_{-2.0}\,,
\eea
thus in a very good agreement with experiment. 
The  small value predicted by the QCD sum rule still needs an explanation.

The value of the coupling $g_{D^\ast D\pi}$ provides also the access 
to the $\widehat g$-coupling which is one of the main parameters of the
approach based on the use of chiral perturbation theory for the 
heavy-light systems. $\widehat g$ is related to $g_{D^\ast D\pi}$ through
\[ \label{eq:2}
g_{D^\ast D \pi}\ =\ \left\{
\begin{array}{ccc} 
{ 2 m_{D^\ast}\  \widehat g_c/ f_\pi }   &   &  \cite{CLEO}\ , \\
&&\\
{ 2 \sqrt{m_D m_{D^\ast}}\ \widehat g_c/ f_\pi }    &   &  \cite{AD}\ , \\
&&\\
{  2 m_D\  \widehat g_c/ f_\pi }   &	&  \cite{casalb}\ ,
\end{array}\right.\;
\]
where we use $f_\pi = 132$~MeV. All the above definitions 
coincide up to $1/m_c$ corrections. Throughout this paper we will use
the most symmetric definition \cite{AD}.
Notice that we assigned a subscript $c$, 
to stress that the $\widehat g$ is not obtained with the infinitely heavy
quark (mesons), but rather from the (not-so-heavy) charmed heavy-light 
mesons.   
By using the definition~\cite{AD},  from our result (\ref{result62}), we obtain
\bea 
\label{eq:1}
\widehat g_c = 0.67 \pm 0.08^{+0.04}_{-0.06}\,.
\eea
We performed the simulations at two values of the lattice spacing 
($\beta =6.0$ and $\beta =6.2$) to study the systematic effects which are
discussed in section 6.  
The simulation was firstly done at $\beta =6.0$ and will be summarized in
section 5. Our main results are given from  the simulations at  $\beta=6.2$ since they
have smaller  ${\cal O}(a)$-effects. This analysis is detailed in sections 2 to 4.

\section{Lattice parametrization and results of the analysis 
       of the two-point functions}

The main results presented in this paper and detailed in this section
are obtained from the 
simulation on a $24^3\times 64$ lattice with periodic boundary 
conditions, at $\beta =6.2$. Our sample contains $100$ independent 
$SU(3)$ gauge configurations produced in the quenched approximation 
({\it i.e.} $n_F=0$). The quark propagators are computed by using the 
following Wilson hopping parameters for light ($q$) and heavy ($Q$) 
quarks:
\bea \label{eq0}
\kappa_q &=& 0.1344_{q_1}\ ,\; 0.1348_{q_2}\ ,\; 0.1351_{q_3}\;;\cr
\kappa_Q &=& 0.1250_{Q_1}\ ,\; 0.1220_{Q_2}\ ,\; 0.1190_{Q_3}\;,
\eea
where we introduced the labels $q_{1-3}$ and $Q_{1-3}$ that will be
used throughout this paper. We have implemented the non-perturbative 
${\cal O}(a)$ improvement of the Wilson fermion action, by setting 
$c_{SW}=1.614$~\cite{alpha1}. 

In this section we consider the standard two-point correlation 
functions

\bea \label{C2} C_{PP}^{(2)}(t_x;\vec p) = \left. \langle  \sum_{\vec x}\, e^{i
\vec p \cdot \vec x} \ P(0) P (x) \rangle \right. ,
\quad  C_{V_\mu
V_\nu}^{(2)}(t_x;\vec p) = \left. \langle  \sum_{\vec x} \, e^{i \vec p \cdot
\vec x} \ V_\mu(0) V_\nu (x) \rangle \right.
\eea
where $t_x>0$ and  where $P\equiv \bar q\, ' \gamma_5 q$ and $V_\mu\equiv \bar q\,'
\gamma_\mu q$ which will be used with heavy-light and light-light quarks. 
We define the constants ${\cal Z}_P$  and ${\cal Z}_V$ so that 

\bea \label{CalZ}
C_{PP}^{(2)}(t_x;\vec p) \simeq  {\cal Z}_P \frac {e^{-E_P t_x}}{2 E_P},\quad 
C_{V_\mu V_\nu}^{(2)}(t_x;\vec p) \simeq {\cal Z}_V \frac {e^{-E_V t_x}}{2 E_V}
(\delta_{\mu\nu} - p_\mu p_\nu/p^2).
\eea
at large $t_x$, where $E_P$ ($E_V$) is the ground state pseudoscalar (vector)
meson energy.  

From the standard study of these two-point light-light correlation functions, we
extracted the masses of pseudoscalar ($a m_P$) and vector mesons ($a m_V$), the decay constant ($a f_P$) and
the (bare improved) light quark mass ($a \rho$) that we obtain by using the axial
 Ward identity, 
 %$\partial_\mu A_\mu^I = 2 i \rho P$.
 $\partial_\mu A_\mu^I = 2  \rho P$. In table~\ref{tab1}, we list our results for both the degenerate 
and non-degenerate combinations of our light quarks.
\begin{table}[ht] 
\begin{center}
\hspace*{-1cm}
\begin{tabular}{|c|c c c c |} 
\hline 
{\phantom{\Huge{l}}}\raisebox{-.2cm}{\phantom{\Huge{j}}}
\hspace*{-7mm}&  $a m_P$  &  $a m_V$ &  $a\rho$  &  $a
f_P^R$ \\  \hline 
{\phantom{\Huge{l}}}\raisebox{-.2cm}{\phantom{\Huge{j}}}  
\hspace*{-7mm} $\kappa_{q_1} - \kappa_{q_1}$ 
& $0.306(1)$ & $0.409(3)$ & $0.0413(4)$ & $0.068(2)$ \\ 
{\phantom{\Huge{l}}}\raisebox{-.2cm}{\phantom{\Huge{j}}}
\hspace*{-7mm} $\kappa_{q_1} - \kappa_{q_2}$  
& $0.284(2)$ & $0.393(4)$ & $0.0354(3)$ & $0.065(2)$ \\ 
{\phantom{\Huge{l}}}\raisebox{-.2cm}{\phantom{\Huge{j}}}
\hspace*{-7mm} $\kappa_{q_1} - \kappa_{q_3}$  
& $0.266(2)$ & $0.381(4)$ & $0.0310(3)$ & $0.063(2)$ \\ 
{\phantom{\Huge{l}}}\raisebox{-.2cm}{\phantom{\Huge{j}}}
\hspace*{-7mm} $\kappa_{q_2} - \kappa_{q_2}$  
& $0.259(2)$ & $0.377(5)$ & $0.0296(3)$ & $0.062(2)$ \\ 
{\phantom{\Huge{l}}}\raisebox{-.2cm}{\phantom{\Huge{j}}}
\hspace*{-7mm} $\kappa_{q_2} - \kappa_{q_3}$  
& $0.240(2)$ & $0.364(5)$ & $0.0252(3)$ & $0.060(2)$ \\ 
{\phantom{\Huge{l}}}\raisebox{-.2cm}{\phantom{\Huge{j}}}
\hspace*{-7mm} $\kappa_{q_3} - \kappa_{q_3}$  
& $0.219(2)$ & $0.350(6)$ & $0.0208(2)$ & $0.058(2)$ \\ \hline 
\end{tabular} 
%\vspace*{.8cm}
\caption{\label{tab1}
\small{\sl Light meson masses, bare 
quark masses and  (renormalized) pseudoscalar decay
constants. The time intervals chosen for the fits are: 
$P: t\in [10,30]$, $V: t\in [11,25]$, and 
$\rho,f_P : t\in [12,29]$.}}
\end{center}
\vspace*{-.3cm}
\end{table}
In the computation of the pseudoscalar decay constant and of the 
bare quark mass  we improved the axial current 
at ${\cal O}(a)$, {\it i.e.}
\bea \label{jed}
 A_\mu^I(x) =  A_\mu(x)\ + \ c_A(g_0^2) \partial_\mu P(x) \;,
\eea
where $c_A = -0.038(4)$, as determined non-perturbatively at $\beta = 6.2$ 
in refs.~\cite{alpha1,LANL,glasgow}. In the computation 
of the renormalized decay constant, the ${\cal O}(a \rho )$ effects are 
eliminated by redefining
\bea \label{fpi}
Z_A^I(g_0^2) = Z_A^{(0)}(g_0^2) \left( 1 + \tilde b_A(g_0^2) a \rho \right)\;,
\eea
where the non-perturbatively estimated constants are 
$Z_{A}^{(0)}=0.81(1)$~\cite{LANL,alpha2,APE},  and $\tilde b_A = 1.19(6)$~\cite{LANL}. 
%%%%%%%%%%%%%%%%%%%%%%%%%%%%%%%%%%%%%%%%%%%%%%%%%%%%%%%%%%%
\begin{figure}[ht]
\vspace*{-.1cm}
\begin{center}
\begin{tabular}{@{\hspace{-0.7cm}}c}
\epsfxsize11.0cm\epsffile{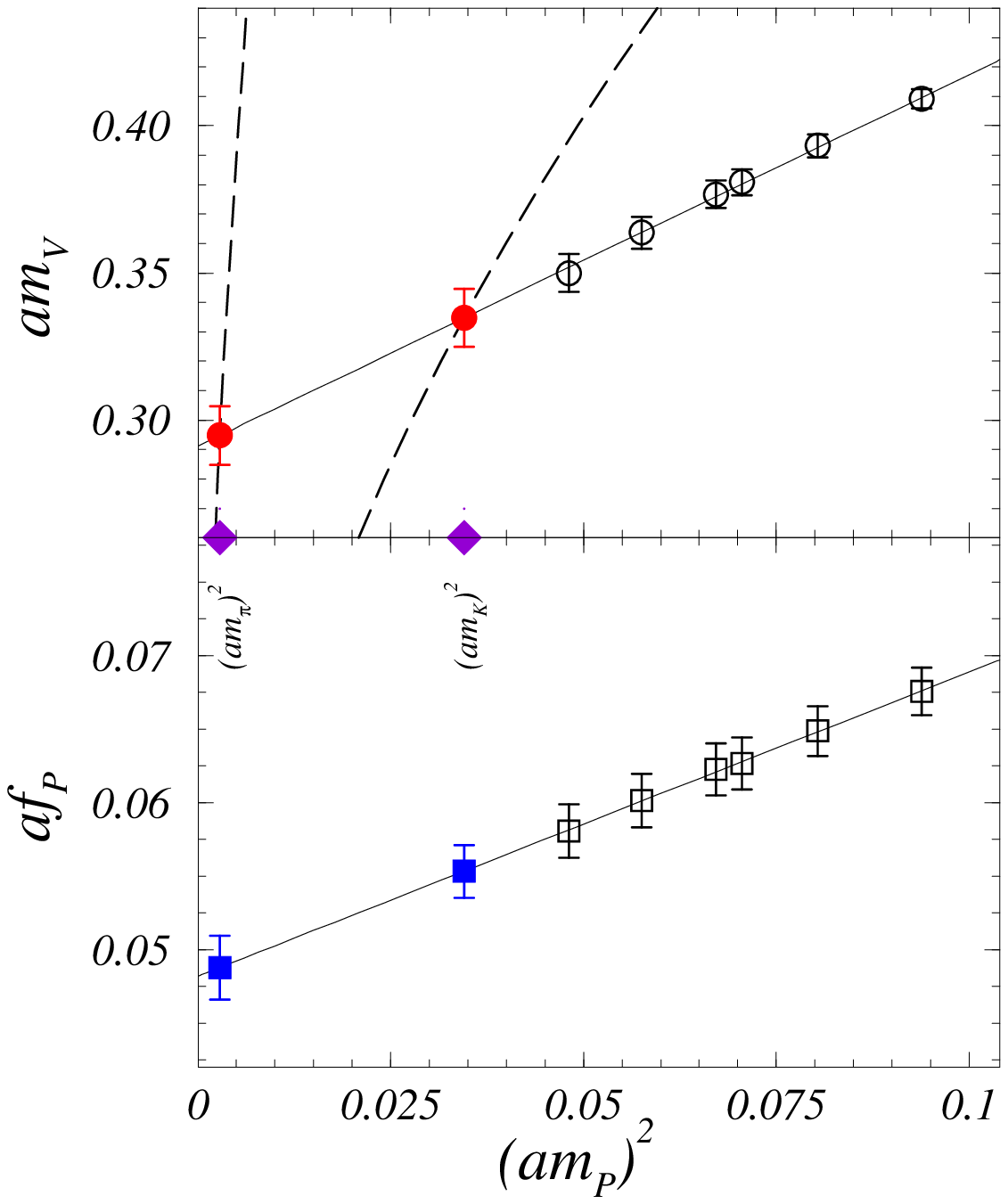}   \\
\end{tabular}
%%%%%%%%%%%%%%%%%%%%%%%%%%%%%%%%%%%%%%%%%%%%%%%%%%%%%%%%%%%%%%%%%%
\caption{\label{fig1}{\small We illustrate the so-called method of lattice
planes \cite{allton} by which we get the value of $a f_{K/\pi}$: In the upper plane, 
the points in which the fit (solid) line to our data crosses the dashed 
lines, corresponding to $am_V = C_{K/\pi} \sqrt{(a m_P)^2}$ with
$C_K=(m_{K^\ast}/m_K)_{phys}$ and $C_\pi=(m_\rho/m_\pi)_{phys}$ respectively,
determine the values of $(a m_{K/\pi})^2$, denoted by diamonds. 
These values are then used to fix $a f_{K/\pi}$ (filled squares) in the 
lower plane, where we fit the pseudoscalar decay constants as
$\alpha + \beta (a m_P)^2$.}}
%%%%%%%%%%%%%%%%%%%%%%%%%%%%%%%%%%%%%%%%%%%%%%%%%%%%%%%%%%%%%%%%%%
\end{center}
\end{figure}
%%%%%%%%%%%%%%%%%%%%%%%%%%%%%%%%%%%%%%%%%%%%%%%%%%%%%%%%%%%

By using the lattice plane method~\cite{allton}, illustrated in 
fig.~\ref{fig1}, we get $a f_\pi = 0.0488(24)$, which after comparison 
with the physical $f_\pi = 0.132$~GeV, leads to the following value 
of the inverse lattice spacing:
\bea \label{am1FPI}
a^{-1}(f_\pi) = 2.71(12)~\gev\;.
\eea
That value is consistent with the one obtained by using 
the $\rho$-meson mass
($a^{-1}(m_\rho) = 2.62(9)$~GeV) and/or the $K^\ast$-meson 
($a^{-1} (m_{K^\ast})= 2.67(8)$~GeV).

As for the heavy-light systems, in what follows, we will need 
the pseudoscalar and vector meson masses, in addition to the constants 
${\cal Z}_{P}  , {\cal Z}_{V} $ (\ref{CalZ}). 
We obtain those quantities by fitting our lattice results for the
correlators~(\ref{C2}) with the mesons at rest ($\vec p=0$), to the forms 
given in eq.(\ref{CalZ}). The results are presented in table~\ref{tab2}.

\begin{table}[ht] 
\begin{center} 
\hspace*{-1cm}
\begin{tabular}{|c|c c c c |} 
\hline 
{\phantom{\Huge{l}}}\raisebox{-.2cm}{\phantom{\Huge{j}}}
\hspace*{-7mm}&  $am_P$  &  $am_V$ &  ${\cal Z}_P$  &  ${\cal Z}_V$ 
 \\  \hline 
{\phantom{\Huge{l}}}\raisebox{-.2cm}{\phantom{\Huge{j}}}
\hspace*{-7mm} $\kappa_{Q_1} - \kappa_{q_1}$ 
& $0.692(2)$ & $0.732(2)$ & $0.0187(9)$ & $0.0068(4)$ \\ 
{\phantom{\Huge{l}}}\raisebox{-.2cm}{\phantom{\Huge{j}}}
\hspace*{-7mm} $\kappa_{Q_1} - \kappa_{q_2}$  
& $0.678(2)$ & $0.719(3)$ & $0.0176(9)$ & $0.0062(5)$ \\ 
{\phantom{\Huge{l}}}\raisebox{-.2cm}{\phantom{\Huge{j}}}
\hspace*{-7mm} $\kappa_{Q_1} - \kappa_{q_3}$  
& $0.669(2)$ & $0.710(4)$ & $0.0171(10)$ & $0.0059(6)$ \\ 
\hline
{\phantom{\Huge{l}}}\raisebox{-.2cm}{\phantom{\Huge{j}}}  
\hspace*{-7mm} $\kappa_{Q_2} - \kappa_{q_1}$ 
& $0.788(2)$ & $0.822(2)$ & $0.0207(10)$ & $0.0080(5)$ \\ 
{\phantom{\Huge{l}}}\raisebox{-.2cm}{\phantom{\Huge{j}}}
\hspace*{-7mm} $\kappa_{Q_2} - \kappa_{q_2}$  
& $0.775(2)$ & $0.809(3)$ & $0.0195(11)$ & $0.0073(6)$ \\ 
{\phantom{\Huge{l}}}\raisebox{-.2cm}{\phantom{\Huge{j}}}
\hspace*{-7mm} $\kappa_{Q_2} - \kappa_{q_3}$  
& $0.766(3)$ & $0.800(4)$ & $0.0189(12)$ & $0.0070(7)$ \\ 
\hline
{\phantom{\Huge{l}}}\raisebox{-.2cm}{\phantom{\Huge{j}}}  
\hspace*{-7mm} $\kappa_{Q_3} - \kappa_{q_1}$ 
& $0.878(2)$ & $0.907(2)$ & $0.0222(11)$ & $0.0091(6)$ \\ 
{\phantom{\Huge{l}}}\raisebox{-.2cm}{\phantom{\Huge{j}}}
\hspace*{-7mm} $\kappa_{Q_3} - \kappa_{q_2}$  
& $0.866(2)$ & $0.894(3)$ & $0.0208(12)$ & $0.0083(7)$ \\ 
{\phantom{\Huge{l}}}\raisebox{-.2cm}{\phantom{\Huge{j}}}
\hspace*{-7mm} $\kappa_{Q_3} - \kappa_{q_3}$  
& $0.857(3)$ & $0.885(4)$ & $0.0203(13)$ & $0.0079(8)$ \\ 
 \hline
\end{tabular} 
%\vspace*{.8cm}
\caption{\label{tab2}
\small{\sl Meson masses and $\cal Z$'s in lattice units extracted from 
our lattice data. Indices in the hopping parameters ($q_{1-3}$ and
$Q_{1-3}$) are specified in eq.~(\ref{eq0}). The time-intervals used for the
fits  are  $P: t\in [15,30]$, $V : t\in [17,28]$.
}}
\end{center}
\vspace*{-.3cm}
\end{table}

\section{Computation of the three-point functions: Basics 
 and Results}
\begin{figure}[ht]
\vspace*{-.1cm}
\begin{center}
\begin{tabular}{@{\hspace{-0.7cm}}c}
\epsfxsize10.0cm\epsffile{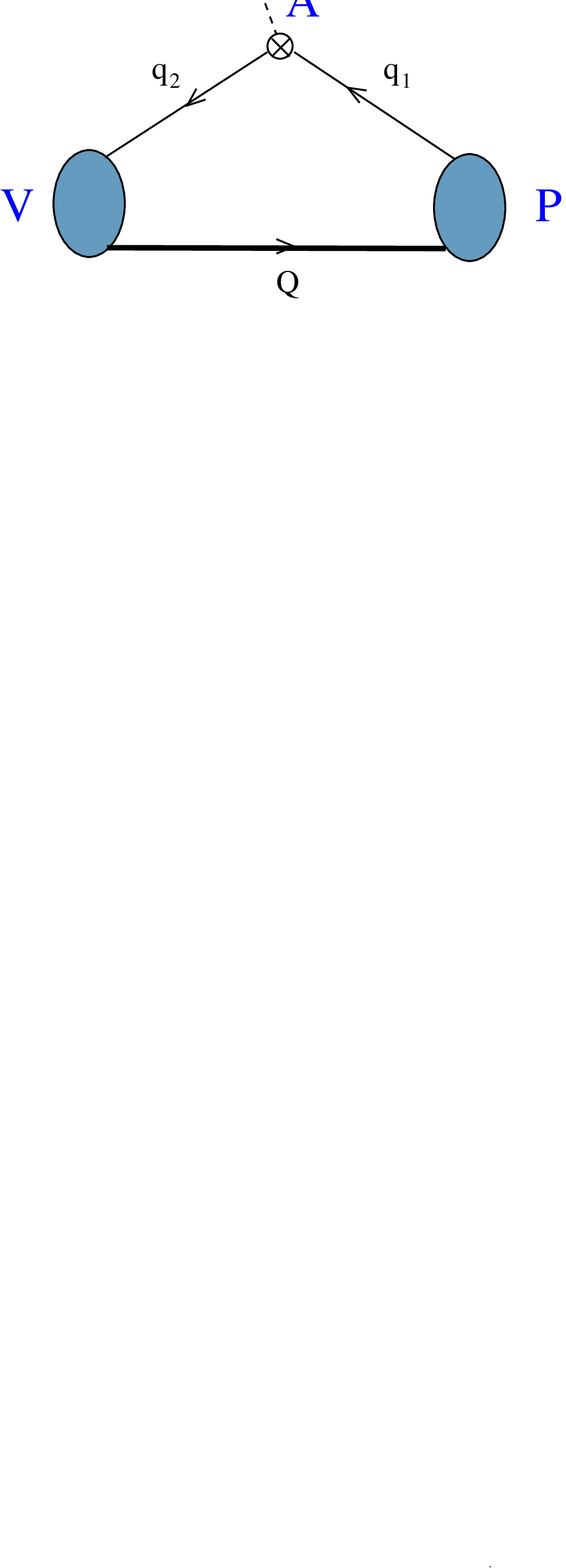}   \\
\end{tabular}
%%%%%%%%%%%%%%%%%%%%%%%%%%%%%%%%%%%%%%%%%%%%%%%%%%%%%%%%%%%%%%%%%%
\caption{\label{fig2}{\small The graph of the three-point function that
we compute in this work.}}
%%%%%%%%%%%%%%%%%%%%%%%%%%%%%%%%%%%%%%%%%%%%%%%%%%%%%%%%%%%%%%%%%%
\end{center}
\end{figure}
%%%%%%%%%%%%%%%%%%%%%%%%%%%%%%%%%%%%%%%%%%%%%%%%%%%%%%%%%%%
In this section we define the quantities that we need to compute and 
explain the strategy that will allow us extracting the coupling $g_{VP \pi}$,
where $P$ and $V$ stand for the heavy-light vector and pseudoscalar mesons 
respectively. We will then present our results obtained for the heavy and 
light quark masses that are directly accessible from our lattice.

\subsection{Theoretical basis}
We start from the computation of the transition matrix element between 
the heavy-light  vector   meson ($V$) and the heavy-light pseudoscalar ($P$), 
mediated by the axial light-light current $A_\mu = \bar q \gamma_\mu 
\gamma_5 q$ (see fig.~\ref{fig2}). It is parametrized as
\bea
\left\langle P(p') \left|{ A^\mu}\right|V(p,\lambda)\right\rangle &=&
2m_{V}{ A_0(q^2)}\frac{\epsilon^{\lambda}\cdot q}{q^2}q^\mu 
+  (m_{P} + m_{V}){ A_1(q^2)}\left[\epsilon^{\lambda\,\mu}-\frac{\epsilon^\lambda\cdot q}{q^2}q^\mu\right]\nonumber\\ 
&& + { A_2(q^2)}\:\frac{\epsilon^\lambda\cdot q}{m_{P}+m_{V}}
\left[p^{\mu}+p^{\prime\,\mu} -\frac{m_{V}^2-m_{P}^2}{q^2}q^{\mu}\right]
\label{ff}
\eea
where $q=p-p'$ ~\footnote{~This definition of form factors is equivalent to the
usual one~: the matrix element  $\left\langle V \left|{
A^\mu}\right|P\right\rangle$ has a ``-" sign  in front of
$A_2$.}. 
The matrix element of the divergence $q_\mu  A^\mu$ is dominated by the pion
 pole for $q^2$ in the vicinity of $m_\pi^2$~:
  \bea
 \langle P(p')|q_\mu A^\mu| V(p,\lambda)\rangle=g_{VP\pi}
 \frac{\epsilon^\lambda(p)\cdot q}{m_\pi^2-q^2}\times f_\pi m_\pi^2+\cdots
 \label{pole}\eea
 From eq. (\ref{pole}), we find for $q^2=0$~:
 \bea
 g_{VP\pi}= \frac{2 m_V A_0(0)}{f_\pi}\ .
 \eea
With our settings, the lattice transfer at $\vec q=\vec 0$  happens to be close
to $q^2=0$ (see table~\ref{tab3}). However, the lattice simulations at $\vec
q=\vec 0$ can only give the form factor $A_{1}$. The other ones, $A_{0,2}$, can
be computed at $\vec q\ne 0$ which in our case is no longer close to $q^2=0$, as
can be seen in table~\ref{tab3}. Since $A_{0}$ has the pion pole, it varies very
fast in the vicinity of $q^2=0$ and thus cannot be directly extrapolated. To
overcome this difficulty, we express $A_0(0)$ in terms of   $A_{1,2}(0)$  which
do not have a pion pole.  This relation can be obtained by using the fact that in
eq.~(\ref{ff}),   the axial current cannot have a singularity  at $q^2=0$.  The
three residues of $1/q^2$-terms must cancel, this leads to:
\bea \label{master}
g_{VP\pi} = \frac 1  {f_\pi}
\biggl[(m_V + m_P) A_1(0) + (m_V - m_P) A_2(0)\biggr]\;.
\eea 
In this relation, one can see that the $A_1$  contribution is dominant. This
dominant contribution is   directly obtained from lattices at $\vec q=\vec 0$,
as  already mentioned. $A_2$ has to be extrapolated from $\vec q\ne\vec 0$ to
$q^2=0$. Its nearest pole is the $a_1$ meson mass. The error generated with this
extrapolation will be discussed in the next subsection. This error is anyhow
harmless since it  applies only to the subdominant contribution ($ \lesssim 5\%$)
to $g_{VP\pi}$.
 
In the heavy quark limit ($m_Q\to \infty$), in which $m_V = m_P$, 
only the form factor $A_{1}(0)$ contributes.
In that limit one recovers the formula used in ref.~\cite{UKQCD} to compute this coupling. 
For heavy quarks close to charm, the corrections in powers of the 
inverse heavy quark mass are expected to be sizable, which is why we
decided to compute $g_{VP\pi}$ with the relativistic (propagating) 
heavy quark.  Notice that 
isospin symmetry relates various charge combinations of the $VP\pi$ couplings:
\bea
g_{VP\pi} \equiv g_{V^+P^0\pi^+} = - \sqrt{2} g_{V^+P^+\pi^0} 
= -g_{V^0P^+\pi^-}\;.
\eea
For simplicity, we define
\bea\label{g12}
G_1(q^2) = {m_V + m_P\over f_\pi} A_1(q^2)\;,\quad
G_2(q^2) = {m_V - m_P\over f_\pi} A_2(q^2)\;,
\eea
and rewrite eq.~(\ref{master}) as
\bea \label{master2}
g_{VP\pi} = G_1(0) \cdot 
\left( 1 + {G_2(0)\over G_1(0)} \right)\;.
\eea
As already mentioned $G_1(0)$ is the dominant contribution 
to $g_{VP\pi}$, the $G_2/G_1$ being a few percent correction to it.

Analogously, the expression for $\widehat g_Q$ at a given heavy quark $Q$ mass
is 
\bea\label{gq}
\widehat g_Q = \widehat g_Q^{\,(0)} \cdot \left( 1 + {G_2(0) \over G_1(0)} \right)\;,
\eea
where we note, according to \cite{AD},  
\bea\label{gq0}
\widehat g_Q^{\,(0)} = { m_V + m_P \over 2 \sqrt{m_V m_P} } \ A_1(0)\;.
\eea

\subsection{Computation of the three point functions}

To access the matrix element~(\ref{ff}) from the lattice, 
we compute the following three-point functions
\bea \label{eq1}
C_{\mu \nu}^{(3)}(0, \vec q, t_x;\vec 0, t_y) = \left. \langle  \sum_{\vec x,\vec y}  \ V_\mu(0) A_\nu (x)
P(y) e^{-i\vec q\cdot \vec x} \rangle \right|_{0< t_x < t_y}\;,
\eea
where the pseudoscalar meson is inserted at rest ($\vec p \ ' = (0,0,0)$) at
a fixed time chosen to be $t_y=31\, a$,  while the current operator
has a momentum $\vec q \in \{ (0,0,0), (1,0,0)\}$ in units of the
elementary momentum  ($2\pi/La\simeq 0.7$~GeV). The vector meson interpolating
field is at the origin $(0,\vec 0)$ of the lattice. 

\begin{figure}[ht]
%\vspace*{-1.cm}
\begin{center}
\begin{tabular}{@{\hspace{-0.7cm}}c}
%\epsfxsize16.0cm\epsffile{FIG_3.eps}   \\
\epsfxsize16.0cm\epsffile{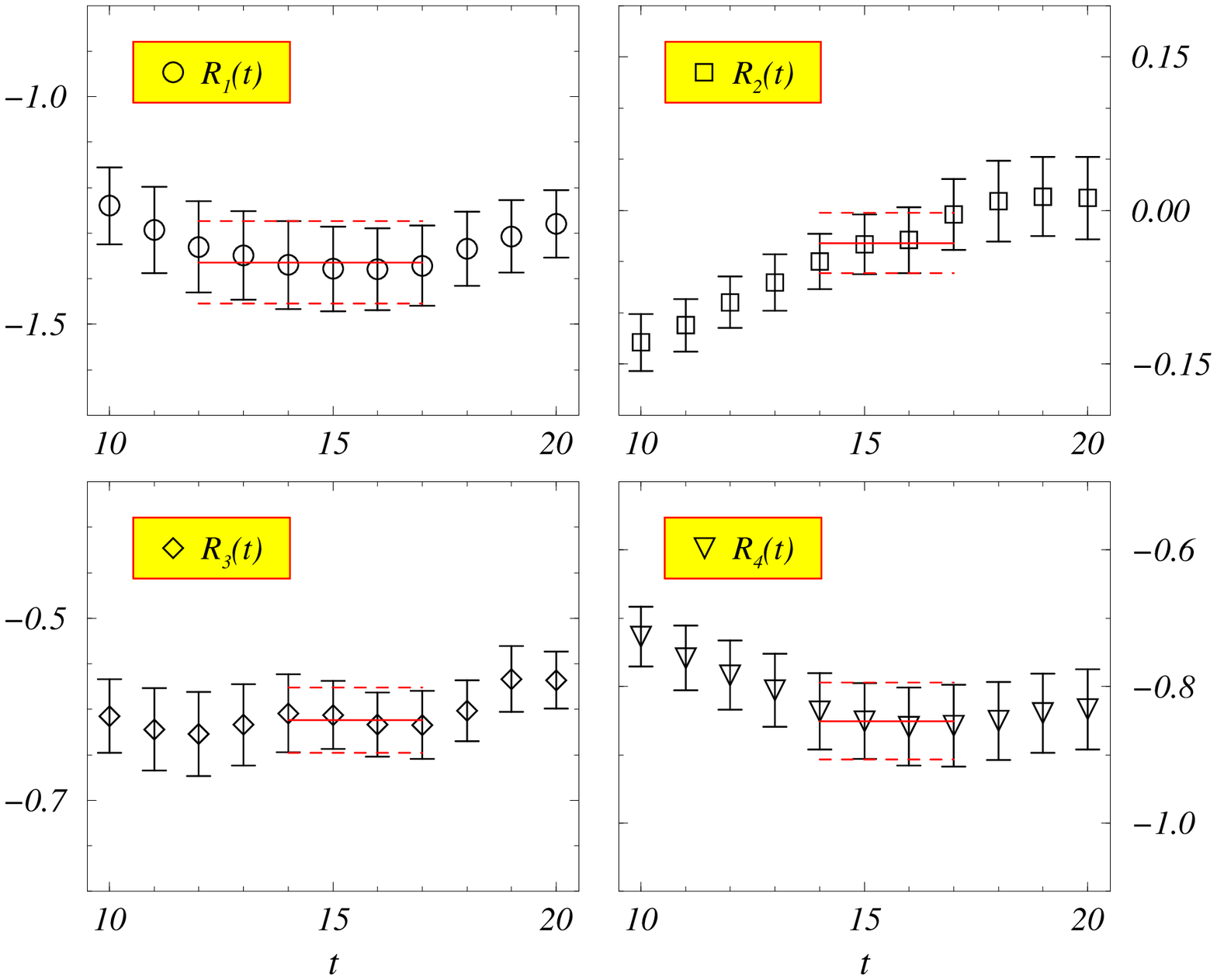}   \\
\end{tabular}
%%%%%%%%%%%%%%%%%%%%%%%%%%%%%%%%%%%%%%%%%%%%%%%%%%%%%%%%%%%%%%%%%%
\caption{\label{fig3}{\small Signals for the ratios $R_{1-4}(t)$ (real part of
$R_1,R_3,R_4$ and imaginary part of $R_2$), defined in
eqs.~(\ref{ratios1}, \ref{ratios2}), as computed on our lattice. 
Illustration is provided for
$\kappa_Q = 0.1220$ and $\kappa_q=0.1348$.   
}}
%%%%%%%%%%%%%%%%%%%%%%%%%%%%%%%%%%%%%%%%%%%%%%%%%%%%%%%%%%%%%%%%%%
\end{center}
\end{figure}

When both mesons are at rest, the only useful ratio is ($t\equiv t_x$)
\bea \label{ratios1}
R_1(t) =  { C^{(3)}_{ii}(t) {\cal Z}_V^{1/2}{\cal Z}_P^{1/2} \over C^{(2)}_{V_i
V_i}(t) C^{(2)}_{P P}(t_y-t)} \;,
\eea
which develops a plateau for $t\in [12,17]$. At that plateau we
extract the matrix element~(\ref{ff}), {\it i.e.} the value of 
the form factor $A_1(q^2)$. To access the ratio $A_2/A_1$, we 
study the ratios with the momentum injection $\vec q =(1,0,0)\times2\pi/L$, namely
\bea \label{ratios2}
R_2(t) &=&  { C^{(3)}_{10}(t;\vec q) {\cal Z}_V^{1/2}{\cal Z}_P^{1/2}\over C^{(2)}_{V_2
V_2}(t;\vec q) C^{(2)}_{P P}(t_y-t)}  \;,\cr
&& \hfill \cr
&& \hfill \cr
R_3(t) &=&  { C^{(3)}_{11}(t;\vec q) {\cal Z}_V^{1/2}{\cal Z}_P^{1/2}\over C^{(2)}_{V_2
V_2}(t;\vec q) C^{(2)}_{P P}(t_y-t)}  \;,\cr
&& \hfill \cr
&& \hfill \cr
R_4(t) &=&  { C^{(3)}_{22}(t;\vec q) {\cal Z}_V^{1/2}{\cal Z}_P^{1/2}\over C^{(2)}_{V_2
V_2}(t;\vec q) C^{(2)}_{P P}(t_y-t)}  \;.
\eea
After inspecting the ratios~(\ref{ratios2}) for all the combinations of heavy and
light quarks, we choose to fit them to a plateau at $t\in [14,17]$.   We
illustrate in fig.~\ref{fig3} the signals for all the four ratios for a given
$\kappa_Q, \kappa_q$. We notice that
the  ratio $R_2(t)$ has a hardly observable plateau,  the value of which is 
compatible with zero. As just stated the ratio  $R_1(t)$ contributes more than
$95\%$ to $g_{VPp}$ and its plateau is  very good.

 %%%%%%%%%%%%%%%%%%%%%%%%%%%%%%%%%%%%%%%%%%%%%%%%%%%%%%%%%%%
%%%%%%%%%%%%%%%%%%%%%%%%%%%%%%%%%%%%%%%%%%%%%%%%%%%%%%%%%%%

Denoting $r_i$ the average value of $R_i$ on the plateaus, 
the form factors $A_{1,2}$ are then given by~:
\bea\label{a1}
&&A_1(\vec q=\vec 0)=-\frac {r_1}{m_V+m_P}\;,\quad 
\quad A_1\left(\vec q=\frac{2\pi}{L}(1,0,0)\right)=-\frac {r_4}{m_V+m_P}\;,\\
&& \hfill \cr
&& \hfill \cr\label{a2}
&&\frac{A_2}{A_1} = 
\frac {(m_P+m_V)^2}{ 2m_P^2 {\vec q}^{\ 2} }\left[
\left({\vec q}^{\ 2} - E_V(E_V-m_P)\right) + 
\frac{m_V^2(E_V-m_P)}{E_V}\frac {r_3}{r_4} + i \frac{m_V^2 q_1}{E_V}
\frac {r_2}{r_4} \right] \, ,
\eea
where $E_V$ is the energy of the vector meson. 
\\ 
 
Our numerical simulations indicate that the ratio ${A_2}/{A_1}$
is positive and of the order of 1.
It results from eq.~(\ref{g12}) that $G_2/G_1\sim (m_V-m_P)/(m_V+m_P)$ which
leads to a small and positive correction to $G_1(0)$ in (\ref{master2}).

In the left part of table~\ref{tab3}, we present our results for the 
form factor $A_1(q^2)$ ($G_1(q^2)$) for all the quark combinations 
and with both mesons at rest, $q^2 = (m_V-m_P)^2$. In this
case, $q^2 \in (0.04, 1.16)\cdot 10^{-2}~\gev^2$.
%; in this case 
%$q^2=(m_V-m_P)^2 \sim 1-3 10^{-2} {\rm GeV}^2$. 
 They are obviously very close to zero and it is reasonable to 
assume that, at such small $q^2$, $G_{1}(q^2) \simeq G_{1}(0)$.

The results for the ratios of the form factors $A_2/A_1$ (or,
equivalently,  $G_2/G_1$) with $\vec q=(1,0,0)\times2\pi/L $  are listed in
the right part of table~\ref{tab3}. They are  obtained from the ratios
$r_{2-4}$ (see eq.~(\ref{a2})). As expected, the ratios of the form factors
$G_2/G_1$ are positive and very small (they never exceed $5\%$). Now this
ratio has to be extrapolated to  $q^2 =0$.  The $A_{1,2}$ form factors may
change significantly but without changing sign since the nearest pole, the
$a_1$-meson, lies at larger $q^2$  ($m_{a_1} \approx
1.2$~\gev~\cite{QCDSF})~\footnote{~This is confirmed by the ratio  $r_4/r_1
= A_1(\vec q = (1,0,0)\times2\pi/L )/A_1(\vec q =\vec 0)\sim 0.6$.}. On the
other hand, the ratio $A_{2}/A_{1}$  is expected to be rather constant
because $A_{1,2}$ have  the same pole factor $1/(1-q^2/m_{a_1}^2)$ which
cancels out in the ratio~\footnote{~$1/(1-q^2/m_{a_1}^2)$ varies of about
$30\%-40\%$ since for $\vec q=(1,0,0)\times2\pi/L $, $q^2 \in (-0.48,
-0.44)~\gev^2$.}. We thus extrapolate $A_{2}/A_{1}$ by keeping this ratio
constant.
  
%%%%%%%%%%%%%%%%%%%%%%%%%%%%%%%%%%%%%%%%%%%%%%%%%%%%%%%%%%%
\begin{table}[t!!] 
\begin{center} 
\begin{tabular}{|c|ccc|ccc|}
\cline{2-7}
\multicolumn{1}{l|}{} &
\multicolumn{3}{c|}{{\phantom{\Huge{l}}}\raisebox{-.1cm}{\phantom{\Huge{j}}}
\underline{$\; \vec q = (0,0,0)\; $}} & 
\multicolumn{3}{c|}{\underline{$\; \vec q = (1,0,0)\times2\pi / L\; $}} \\
\hline 
{\phantom{\Huge{l}}}\raisebox{-.2cm}{\phantom{\Huge{j}}}  
\hspace*{-5mm}&  $ q^2\times 10^{-2}~\gev^2 $  & $A_1(q^2)$  & $G_1(q^2)$  &
$ q^2 \times 10^{-2}~\gev^2$  & $A_2/A_1$  & $G_2/G_1$ 
  \\   \hline \hline 
{\phantom{\Huge{l}}}\raisebox{-.2cm}{\phantom{\Huge{j}}}  
\hspace*{-5mm} $\kappa_{Q_1} - \kappa_{q_1}- \kappa_{q_1}$ & 
   $1.17(17)$   &   $0.71(4)$	& $14.9\pm 0.9$  &   
   $-44.8(3.9)$   &   $0.85(14)$	& $0.024(4)$ \\
{\phantom{\Huge{l}}}\raisebox{-.2cm}{\phantom{\Huge{j}}}
\hspace*{-5mm} $\kappa_{Q_1} - \kappa_{q_1}- \kappa_{q_2}$ & 
   $0.52(12)$   &   $0.68(6)$	& $14.8 \pm 1.2$  &   
   $-46.3(4.0)$   &   $0.74(12)$	& $0.014(2)$   \\
{\phantom{\Huge{l}}}\raisebox{-.2cm}{\phantom{\Huge{j}}}
\hspace*{-5mm} $\kappa_{Q_1} - \kappa_{q_1}- \kappa_{q_3}$ & 
   $0.23(10)$  &   $0.68(6)$	& $15.1 \pm 1.3$  &  
   $-47.1(4.1)$  &   $0.70(19)$	& $0.009(2)$   \\
{\phantom{\Huge{l}}}\raisebox{-.2cm}{\phantom{\Huge{j}}}
\hspace*{-5mm} $\kappa_{Q_1} - \kappa_{q_2}- \kappa_{q_2}$ & 
   $1.20(22)$  &   $0.69(5)$	& $15.5 \pm 1.2$  &  
   $-44.7(3.9)$  &   $0.93(22)$	& $0.027(6)$  \\
{\phantom{\Huge{l}}}\raisebox{-.2cm}{\phantom{\Huge{j}}}
\hspace*{-5mm} $\kappa_{Q_1} - \kappa_{q_2}- \kappa_{q_3}$ & 
   $0.73(21)$  &   $0.67(7)$	& $15.5 \pm 1.5$  &  
   $-45.7(3.9)$  &   $0.85(25)$	& $0.019(5)$   \\
{\phantom{\Huge{l}}}\raisebox{-.2cm}{\phantom{\Huge{j}}}
\hspace*{-5mm} $\kappa_{Q_1} - \kappa_{q_3}- \kappa_{q_3}$ & 
   $1.24(30)$  &   $0.68(5)$	& $16.0 \pm 1.5$  & 
   $-44.5(3.8)$  &   $1.13(35)$	& $0.034 (9)$  \\
 \hline
 
{\phantom{\Huge{l}}}\raisebox{-.2cm}{\phantom{\Huge{j}}}  
\hspace*{-5mm} $\kappa_{Q_2} - \kappa_{q_1}- \kappa_{q_1}$ & 
   $0.82(12)$   &   $0.71(4)$	& $16.9\pm 1.1$  &   
    $-46.1(4.1)$   &   $0.89(14)$	& $0.018(3)$  \\
{\phantom{\Huge{l}}}\raisebox{-.2cm}{\phantom{\Huge{j}}}
\hspace*{-5mm} $\kappa_{Q_2} - \kappa_{q_1}- \kappa_{q_2}$ & 
   $0.31(9)$   &   $0.68(6)$	& $16.7 \pm 1.4$  &  
   $-47.3(4.2)$   &   $0.77(14)$	& $0.010(2)$  \\
{\phantom{\Huge{l}}}\raisebox{-.2cm}{\phantom{\Huge{j}}}
\hspace*{-5mm} $\kappa_{Q_2} - \kappa_{q_1}- \kappa_{q_3}$ & 
   $0.11(7)$  &   $0.68(6)$	& $17.1 \pm 1.5$  &  
    $-48.0(4.2)$  &   $0.73(22)$	& $0.006(2)$  \\
{\phantom{\Huge{l}}}\raisebox{-.2cm}{\phantom{\Huge{j}}}
\hspace*{-5mm} $\kappa_{Q_2} - \kappa_{q_2}- \kappa_{q_2}$ & 
   $0.84(17)$  &   $0.69(5)$	& $17.5 \pm 1.5 $  & 
   $-46.0(4.0)$  &   $1.19(22)$	& $0.025(4)$   \\
{\phantom{\Huge{l}}}\raisebox{-.2cm}{\phantom{\Huge{j}}}
\hspace*{-5mm} $\kappa_{Q_2} - \kappa_{q_2}- \kappa_{q_3}$ & 
   $0.47(16)$  &   $0.68(7)$	& $17.7 \pm 1.7$  &  
   $-46.9(4.1)$  &   $0.92(29)$	& $0.015(4)$  \\
{\phantom{\Huge{l}}}\raisebox{-.2cm}{\phantom{\Huge{j}}}
\hspace*{-5mm} $\kappa_{Q_2} - \kappa_{q_3}- \kappa_{q_3}$ & 
   $0.85(24)$  &   $0.68(6)$	& $18.3 \pm 1.8$  & 
   $-45.9(3.9)$  &   $1.27(38)$	& $0.028(7)$  \\
 \hline
 
{\phantom{\Huge{l}}}\raisebox{-.2cm}{\phantom{\Huge{j}}}  
\hspace*{-5mm} $\kappa_{Q_3} - \kappa_{q_1}- \kappa_{q_1}$ & 
   $0.60(9)$   &   $0.71(4)$	& $18.9\pm 1.3$  &   
   $-46.9(4.2)$   &   $0.93(14)$	& $0.015(2)$   \\
{\phantom{\Huge{l}}}\raisebox{-.2cm}{\phantom{\Huge{j}}}
\hspace*{-5mm} $\kappa_{Q_3} - \kappa_{q_1}- \kappa_{q_2}$ & 
   $0.19(6)$   &   $0.68(7)$	& $18.5 \pm 1.7$  &   
   $-48.0(4.2)$   &   $0.80(15)$	& $0.007(1)$  \\
{\phantom{\Huge{l}}}\raisebox{-.2cm}{\phantom{\Huge{j}}}
\hspace*{-5mm} $\kappa_{Q_3} - \kappa_{q_1}- \kappa_{q_3}$ & 
   $0.04(5)$  &   $0.68(7)$	& $19.1 \pm 1.8$  &  
   $-48.7(4.3)$  &   $0.78(25)$	& $0.003(1)$  \\
{\phantom{\Huge{l}}}\raisebox{-.2cm}{\phantom{\Huge{j}}}
\hspace*{-5mm} $\kappa_{Q_3} - \kappa_{q_2}- \kappa_{q_2}$ & 
   $0.61(14)$  &   $0.70(5)$	& $19.6 \pm 1.7$  &  
   $-46.9(4.1)$  &   $1.08(23)$	& $0.018(4)$  \\
{\phantom{\Huge{l}}}\raisebox{-.2cm}{\phantom{\Huge{j}}}
\hspace*{-5mm} $\kappa_{Q_3} - \kappa_{q_2}- \kappa_{q_3}$ & 
   $0.62(20)$  &   $0.68(7)$	& $19.9 \pm 2.1$  &  
   $-47.7(4.2)$  &   $1.00(32)$	& $0.012(3)$  \\
{\phantom{\Huge{l}}}\raisebox{-.2cm}{\phantom{\Huge{j}}}
\hspace*{-5mm} $\kappa_{Q_3} - \kappa_{q_3}- \kappa_{q_3}$ & 
   $0.25(8)$  &   $0.68(6)$	& $20.5 \pm 2.2$  &  
   $-46.9(4.1)$  &   $1.43(41)$	& $0.024(6)$  \\
\hline
\end{tabular} 
%\vspace*{.8cm}
\caption{\label{tab3}
\small{\sl  $A_1, G_1, A_2/A_1, G_2/G_1$
at $\beta=6.2$ for several values of $q^2$ in the physical units 
($\gev^2$) as obtained by using $a^{-1}=2.71(12)$ GeV (from eq.~(\ref{am1FPI})). 
}}
\end{center}
%\vspace*{-.3cm}
\end{table}
%%%%%%%%%%%%%%%%%%%%%%%%%%%%%%%%%%%%%%%%%%%%%%%%%%%%%%%%%%%

\subsection{Comment about the effect of improvement on the form factors}

As a side remark, we comment on the effect of the improvement of the bare 
axial current (see eq.~(\ref{jed})) on the form factors $A_{1,2,0}(q^2)$. 
The divergence of the current in the matrix element~(\ref{ff}) 
leads to
\bea \label{matP}
\langle P(p^\prime)\vert \bar q \gamma_5 q \vert V(p, \lambda)\rangle =
 { 2 m_V} {(e_\lambda\cdot q) \over
2 \rho_q} \, A_0(q^2)  \;,
\eea
where we used eq.~(\ref{ff}) and the axial Ward identity,
% $\partial_\mu A_\mu^I = 2 i \rho_q P$. 
 $\partial_\mu A_\mu^I = 2  \rho_q P$. Therefore the improvement will only affect the form factor 
$A_0(q^2)$, but not the other two ($A_{1,2}(q^2)$). 
In other words, 
\bea \label{imprA0}
A_{1,2}(q^2) &\to& A_{1,2}^I(q^2)= A_{1,2}(q^2) \;,\cr
A_0(q^2) &\to& A_0^I(q^2)= \left( 1 - c_A {q^2\over 2 \rho_q}\right) A_0(q^2)\;.
\eea
As our method to compute $g_{D^\ast D\pi}$ relies on the computation
     of the form factors $A_1$ and $A_2$ only, our
     result does not depend at all on the improvement of the bare axial
     current.

\section{Chiral extrapolations and the heavy quark interpolation}

Now it is a simple matter to combine our results from table~\ref{tab3} 
in the way indicated in eq.~(\ref{master2}), and to compute the values of the 
couplings $g_{VPp}$~\footnote{~The lower subscript ``p"  labels the 
light pseudoscalar meson.}. Those numbers should now be extrapolated to the 
 to the physical pion mass, to get $g_{VP\pi}$ 
for each of our heavy quarks. Finally this is to be followed by interpolation 
in the heavy meson masses  to the ones corresponding to $D$ and $D^\ast$.

\subsection{Chiral extrapolations}

Our light pseudoscalar mesons (``pions'') are rather heavy and we need 
to extrapolate to the physical pion mass~\footnote{~In physical units, our light 
pseudoscalar mesons are in the range $m_p \in (0.5, 0.8)$~GeV.}, 
$m_\pi = 0.14$~GeV. From the lattice planes method \cite{allton}, 
the physical pion 
mass in lattice units is $a m_\pi = 0.053(2)$. To obtain $g_{VP\pi}$, we can  
fit  the $g_{VPp}$-dependence on the quark mass with three formulae:
\begin{itemize}
\item[--] {\sf Linear extrapolation:} by fitting our data to 
\bea \label{fit1}
g_{VPp} = a_0 + a_1 (am_p)^2\;,
\eea
where the values of $(am_p)$ are listed in the first column of table~\ref{tab1}.
This fit allows to fix the parameters $a_{0,1}$ and thus to extract 
$g_{VP\pi}$ by either simply reading off  $a_0$ (chiral limit), 
or by combining $a_{0,1}$ with $a m_\pi$. We will give results by 
opting for the  latter choice.
\item[--] {\sf Quadratic extrapolation:}  we also  attempt the quadratic fit~: 
\bea \label{fit2}
g_{VPp} = b_0 + b_1 (am_p)^2 + b_2 (am_p)^4\;.
\eea
\item[--] {\sf ``Chiral Log'' extrapolation:} One can also fit to a form 
motivated by the chiral perturbative expansion for the heavy-light systems 
(for a review, see ref.~\cite{casalb}). In particular, the 
coupling of the pion to the heavy meson doublet receives logarithmic 
corrections~\cite{falk}~:
\bea \label{fit3}
g_{VPp} = c_0 + c_1 (am_p)^2 + c_2 (am_p)^2 \log((a m_p)^2)\;.
\eea 
\end{itemize}

\begin{figure}[ht]
\vspace*{-.1cm}
\begin{center}
\begin{tabular}{@{\hspace{-0.7cm}}c}
\epsfxsize11.0cm\epsffile{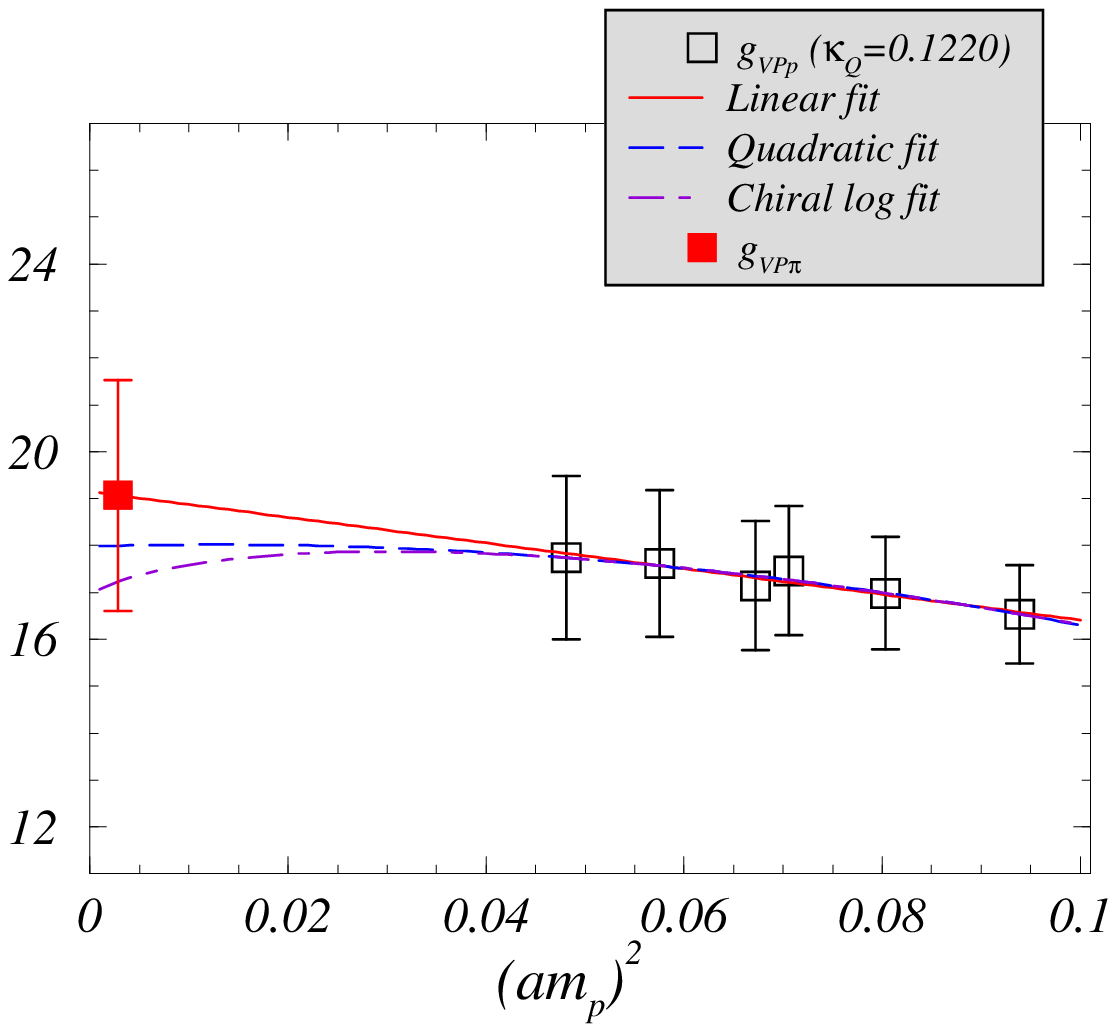}   \\
\end{tabular}
%%%%%%%%%%%%%%%%%%%%%%%%%%%%%%%%%%%%%%%%%%%%%%%%%%%%%%%%%%%%%%%%%%
\caption{\label{fig4}{\small Chiral extrapolation of $g_{VPp}$ without 
the $G_2/G_1$ corrections for a
fixed heavy quark at $\beta=6.2$. The curves obtained from the fit to 
eqs.~(\ref{fit1},\ref{fit2},\ref{fit3}) are displayed. The filled square 
point with its error bars corresponds to the 
linear extrapolation. The complete list of result can be found in
table~\ref{tab4}.   
}}
%%%%%%%%%%%%%%%%%%%%%%%%%%%%%%%%%%%%%%%%%%%%%%%%%%%%%%%%%%%%%%%%%%
\end{center}
\end{figure}

The results of all three fits are presented in table~\ref{tab4} and illustrated
in fig.~\ref{fig4}. The quadratic and logarithmic extrapolations 
are rather unstable i.e. significantly dependent on small changes in the analysis 
procedure, because we extrapolate from too heavy quark masses. We present them
mainly as an estimate of the systematic error.

%%%%%%%%%%%%%%%%%%%%%%%%%%%%%%%%%%%%%%%%%%%%%%%%%%%%%%%%%%%
%%%%%%%%%%%%%%%%%%%%%%%%%%%%%%%%%%%%%%%%%%%%%%%%%%%%%%%%%%%

%%%%%%%%%%%%%%%%%%%%%%%%%%%%%%%%%%%%%%%%%%%%%%%%%%%%%%%%%%%
\begin{table}[ht] 
\begin{center}
\hspace*{-1cm}
\begin{tabular}{|c|c c c c c c |} 
\hline 
{\phantom{\Huge{l}}}\raisebox{-.2cm}{\phantom{\Huge{j}}}
\hspace*{-7mm}{$\kappa_Q$}&  $a m_P$  & $a m_V$  &  $G_1(0)^{(lin.)}$ & 
$G_1(0)^{(quad.)}$  &  $G_1(0)^{(log.)}$ &  $g_{VP\pi}^{(lin.)}$\\  \hline 
{\phantom{\Huge{l}}}\raisebox{-.2cm}{\phantom{\Huge{j}}}  
\hspace*{-7mm} $Q_1 : 0.1250$ 
& $0.645(3)$ & $0.688(6)$ & $16.8\pm 1.9$& $15.6\pm 2.4$ & $15.0\pm 3.0$  &
$17.7 \pm 2.2$\\ 
{\phantom{\Huge{l}}}\raisebox{-.2cm}{\phantom{\Huge{j}}}
\hspace*{-7mm} $Q_2 : 0.1220$ 
& $0.744(4)$ & $0.779(6)$ & $19.3\pm 2.4$ & $18.0\pm 3.3$ & $17.2\pm 4.2$  &
$20.1 \pm 2.7$\\ 
{\phantom{\Huge{l}}}\raisebox{-.2cm}{\phantom{\Huge{j}}}
\hspace*{-7mm} $Q_3 : 0.1190$  
& $0.836(4)$ & $0.865(7)$ & $21.7\pm 3.0$ & $20.3\pm 4.3$ & $19.3\pm 5.6$   &
$22.6 \pm 3.3$\\ \hline 
\end{tabular} 
%\vspace*{.8cm}
\caption{\label{tab4}
\small{\sl For each heavy quark directly simulated on the lattice we
show the values of the heavy-light meson masses for which the light quark
is (linearly) extrapolated to the $u/d$-quark mass. 
We also list the values of the corresponding  $G_1(0)$ ({\it i.e.} the $g_{VP\pi}$ coupling 
without the $G_2/G_1$ correction) by using 
eqs.~(\ref{fit1}, \ref{fit2}, \ref{fit3}), and in the last column,
the $g_{VP\pi}$ linearly extrapolated  including the $G_2/G_1$ 
correction (see eq.~(\ref{master2})).}}
\end{center}
\vspace*{-.3cm}
\end{table}

\begin{table}[ht] 
\begin{center}
\hspace*{-1cm}
\begin{tabular}{|c| c c |} 
\hline 
{\phantom{\Huge{l}}}\raisebox{-.2cm}{\phantom{\Huge{j}}}
\hspace*{-7mm}{$\kappa_Q$}&    $\widehat g_Q^{\,(0,\,lin.)}$ 
 &  $\widehat g_Q^{\,(lin.)}$ \\  \hline 
{\phantom{\Huge{l}}}\raisebox{-.2cm}{\phantom{\Huge{j}}}  
\hspace*{-7mm} $Q_1 : 0.1250$ & $0.636(65)$  & $0.669(72)$ \\ 
{\phantom{\Huge{l}}}\raisebox{-.2cm}{\phantom{\Huge{j}}}
\hspace*{-7mm} $Q_2 : 0.1220$ & $0.641(76)$  & $0.668(79)$ \\ 
{\phantom{\Huge{l}}}\raisebox{-.2cm}{\phantom{\Huge{j}}}
\hspace*{-7mm} $Q_3 : 0.1190$  & $0.648(86)$  &  $0.673(88)$ \\ \hline 
\end{tabular} 
%\vspace*{.8cm}
\caption{\label{tab4-bis}
\small{\sl For each heavy quark we
show the values of $\widehat g_Q^{\,(0)}$  for which the light quark
is linearly  extrapolated to the $u/d$-quark mass and 
the $\widehat g_Q$ linearly extrapolated including the $G_2(0)/G_1(0)$
correction (see eq.(\ref{gq})). 
}}
\end{center}
\vspace*{-.3cm}
\end{table}

\subsection{Chiral extrapolations with two different light quarks}

We have computed the three-point Green functions with two different  light quark
masses $m_1, m_2$ for the quarks $q_1, q_2$ shown in fig~\ref{fig2}. The form
factors extracted from the latter depend, in general, both on  $m_2~+~m_1$ and
$m_2-m_1$ what makes the chiral extrapolation rather tricky. Luckily, however, it
can be shown \footnote{~Using the heavy quark symmetry and the hermiticity
of $A_\mu$, one derives that\\
$<V(m_2)|A_\mu|P(m_1)> = <P(m_2)|A_\mu|V(m_1)> = <V(m_1)|A_\mu|P(m_2)>$.} 
that, in the heavy quark limit,  the dominant contribution to 
$g_{VP\pi}$, {\it i.e.} $G_1(0)$, only depends on  $m_2+m_1$. In this limit, we
can use  data with $(m_1 \ne m_2)$ to perform our chiral extrapolation as a
function of $m_2~+~m_1$ (or similarly, as a function of $(m_p)^2$). We have
compared  the results of the linear extrapolation using this method with the one
which uses Green functions with only $m_1~=~m_2$. The extrapolated results agree
within $2 \%$ while the statistical error is typically $\sim 15 \%$.  
Considering  six $(m_1, m_2)$ couples allows to estimate the systematic
error due to the chiral extrapolation by comparing the linear, quadratic and
logarithmic extrapolations as explained in the preceding subsection. As can be
seen in table~\ref{tab4}, $G_1(0)$ was computed in this way.

The quark mass dependence of the corrective term $G_2(0)/G_1(0)$ in
eqs.~(\ref{master2},\ref{gq}) is dominated by the factor $m_V-m_P$ in
eq.~(\ref{g12}). In the infinite mass limit, it is known that 
$m_V-m_P \propto m_2-m_1 + {\cal O}(1/m_Q)$. 
The corrective term $G_2(0) /G_1(0)$ is then dominantly proportional
to $m_2-m_1$~\footnote{~We have indeed checked that the slope of 
$G_2(0)/G_1(0)$
as a function of $m_2-m_1$ is four times larger than the one as a function of
$m_1+m_2$.} and the contributions with $m_1\ne m_2$ are useless for the
chiral extrapolation. Therefore, we have only considered the case $m_1=m_2$ when including the correction $G_2(0)/G_1(0)$.

\subsection{Interpolation to the charm sector}

Finally, we need to reach the mass of the charm quark. To that end we
will use the values of the spin-averaged masses of the heavy-light mesons
\bea
 \overline{ m }_H = {3 m_V + m_P \over 4}\;.
\eea
When converted to the physical units by means of $a^{-1}(f_\pi)$ (given in 
eq.~(\ref{am1FPI})), we have
\bea
\overline{ m }_H \in \left\{\ 1.83(9),\, 2.08(10),\, 2.32(11)\  \right\}\ \gev\;.
\eea
and $\overline{ m }_D = 1.974$~GeV is within this range. 
The coupling $\widehat g_Q$ that we already
mentioned in the introduction~:
\bea
g_{VP\pi}\ =\ { 2 \sqrt{m_P m_V}\over f_\pi}\ \widehat g_Q  \  .
\eea
is the proper parameter to be used for an interpolation in the heavy quark mass. 
The heavy quark symmetry suggests us to fit our data to the following 
forms
\bea \label{fitts}
{g_{VP\pi}\over \overline{ m }_H} =   a_1 + { a_2 \over \overline{ m }_H } \;\quad {\rm and}\quad 
\widehat g_{Q}  = b_1 + { b_2 \over \overline{ m }_H }\;.
\eea
Using these forms we now interpolate to the charm region each of our three sets
of 
chirally extrapolated data from the previous subsection. The results are 
presented in table~\ref{tab5}.
\begin{table}[ht] 
\begin{center}
\hspace*{-1cm}
\begin{tabular}{|c|c c c|} 
\hline 
{\phantom{\Huge{l}}}\raisebox{-.2cm}{\phantom{\Huge{j}}}
\hspace*{-7mm}{$\chi$-extrap.}&  $g_{D^\ast D\pi}$  & $\widehat g_{c}$  &  
$\widehat g_{\infty}$  \\  \hline 
{\phantom{\Huge{l}}}\raisebox{-.2cm}{\phantom{\Huge{j}}}  
\hspace*{-7mm} linear 
& $18.82\pm 2.34$ & $0.669(75)$ & $0.69(18)$  \\ 
{\phantom{\Huge{l}}}\raisebox{-.2cm}{\phantom{\Huge{j}}}
\hspace*{-7mm} quadratic 
& $17.11\pm 3.10$ & $0.574(88)$ & $0.62(27)$   \\ 
{\phantom{\Huge{l}}}\raisebox{-.2cm}{\phantom{\Huge{j}}}
\hspace*{-7mm} $\chi$-log  
& $16.39\pm 4.02$ & $0.551(118)$ & $0.57(37)$   \\ \hline 
\end{tabular} 
%\vspace*{.8cm}
\caption{\label{tab5}
\small{\sl Results of the linear interpolation of the form~(\ref{fitts}) 
to $D$-$D^\ast$ mesons at $\beta=6.2$, for each of the three chiral extrapolations discussed
in the text. Note that only the linearly extrapolated result incorporates the 
$G_2/G_1$ correction. The other two are uncorrected.}}
\end{center}
\vspace*{-.3cm}
\end{table}
\begin{figure}[ht]
\vspace*{-.1cm}
\begin{center}
\begin{tabular}{@{\hspace{-0.7cm}}c}
\epsfxsize17.0cm\epsffile{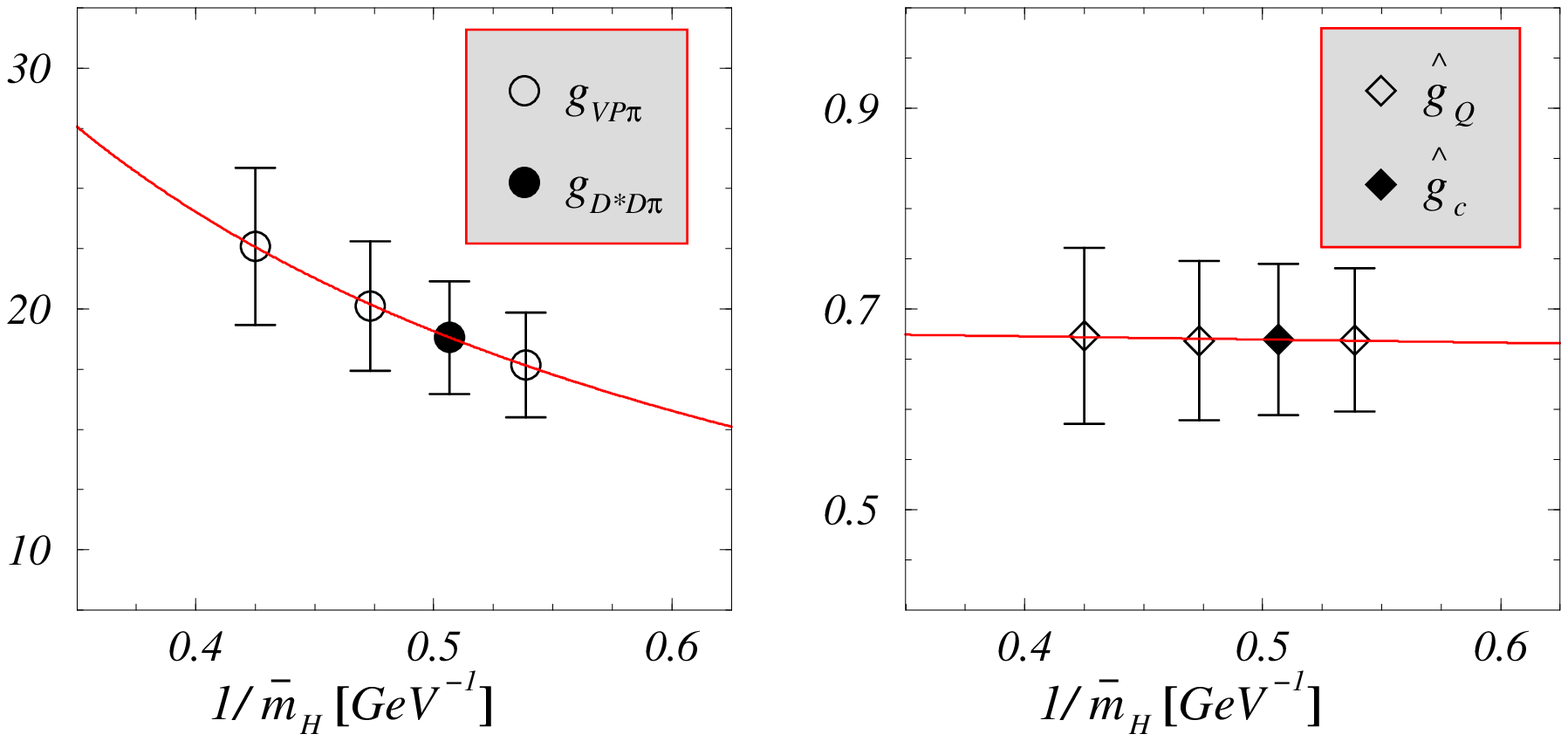}   \\
\end{tabular}
%%%%%%%%%%%%%%%%%%%%%%%%%%%%%%%%%%%%%%%%%%%%%%%%%%%%%%%%%%%%%%%%%%
\caption{\label{fig5}{\small \sl Fit of our data (empty symbols) for $g_{VP\pi}$ and $\widehat g_{Q}$,
at $\beta=6.2$, to  the forms~(\ref{fitts}). The results of interpolation to the  $\bar D$-meson is
denoted by the filled symbols.}}
%%%%%%%%%%%%%%%%%%%%%%%%%%%%%%%%%%%%%%%%%%%%%%%%%%%%%%%%%%%%%%%%%%
\end{center}
\end{figure}
%%%%%%%%%%%%%%%%%%%%%%%%%%%%%%%%%%%%%%%%%%%%%%%%%%%%%%%%%%%

Illustration of that interpolation for the case of the linear chiral 
extrapolation is provided in fig.~\ref{fig5}.
Note that our results indicate that the slope in $1/m_H$ 
for the coupling $\widehat g_Q$ is small and negative. Assuming that
the linear dependence in $1/m_H$ holds all the way to $1/m_H \to 0$, we 
obtain that $\widehat g_\infty$ is not more than $15$~\% larger than 
$\widehat g_c$ (numerical results for $\widehat g_\infty$ are also 
given in table~\ref{tab5}).

%%%%%%%%%%%%%%%%%%%%%%%%%%%%%%%%%%%%%%%%%%%%%%%%%%%%%%%%%%%
\section{Results at $\beta=6.0$}
%%%%%%%%%%%%%%%%%%%%%%%%%%%%%%%%%%%%%%%%%%%%%%%%%%%%%%%%%%%

In this section we briefly summarize the results obtained at $\beta=6.0$. The
analysis follows the lines presented in the previous sections. We have run  in
a $16^3\times 64$ volume over 100 configurations  with the following set of light
quarks Wilson hopping parameters: $\kappa_q~=~0.1339,~0.1342,~0.1344,~0.1346$.  
We will only show results for one heavy quark, the closest to the physical charm,
$\kappa_Q=0.1190$ ($m_P=1.77(11)~\gev $), for several values of the light quark
mass (see table~\ref{tab7}).  For $\beta=6.0$, the ratio $G_2(0)/G_1(0)$  has
about $100 \%$ error, and we  prefer to give the uncorrected result $g_Q^{\,(0)}$
(see eq.~(\ref{gq})), in table~\ref{tab7}.

%%%%%%%%%%%%%%%%%%%%%%%%%%%%%%%%%%%%%%%%%%%%%%%%%%%%%%%%%%%%%%%%%%
\begin{figure}[ht]
\vspace*{-.1cm}
\begin{center}
%%%%%%%%%%%%%%%%%%%%%%%%%%%%%%%%%%%%%%%%%%%%%%%%%%%%%%%%%%%%%%%%%%
\begin{tabular}{@{\hspace{-0.7cm}}c}
\epsfxsize11.0cm\epsffile{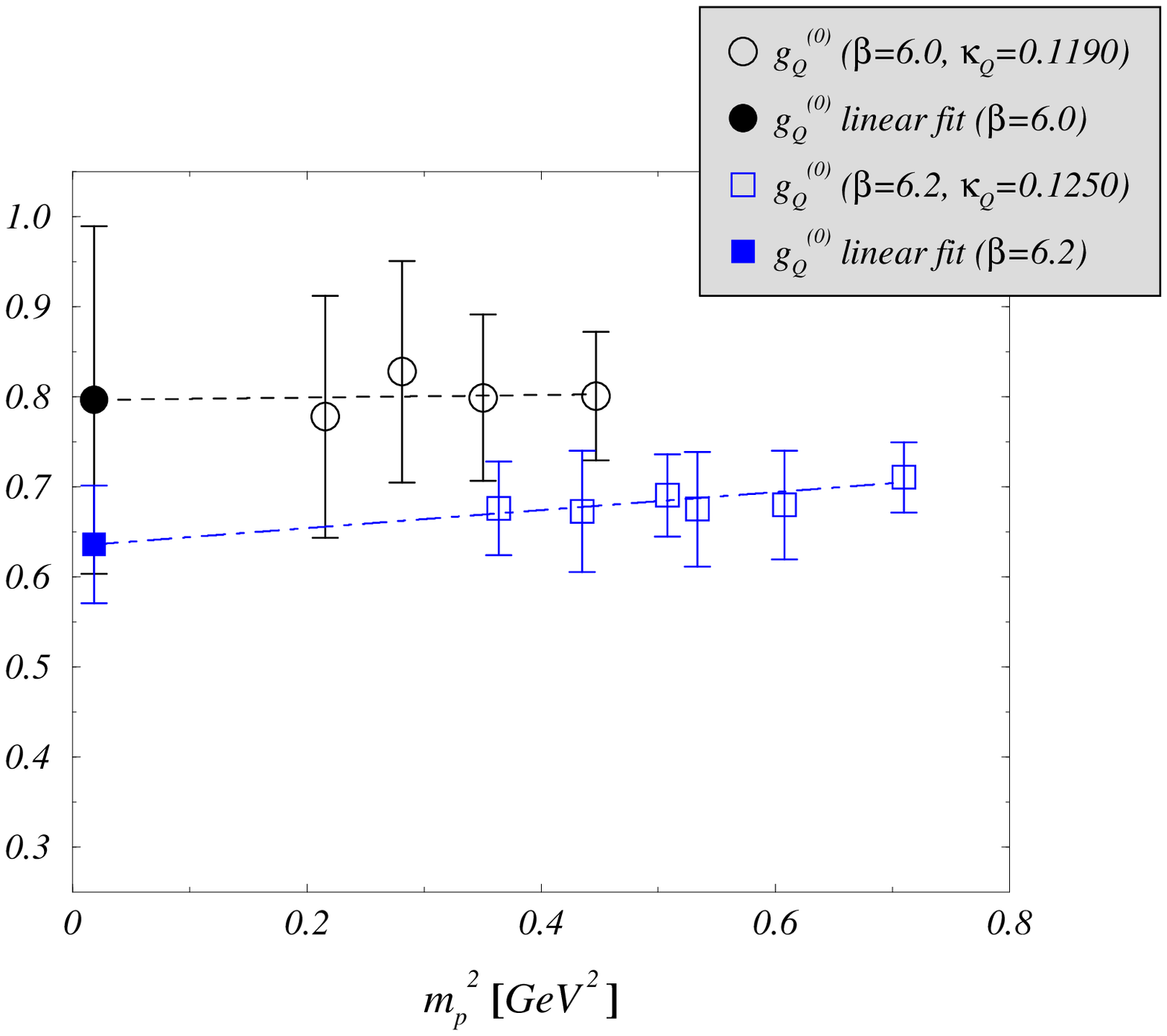}  \\
\end{tabular}
%%%%%%%%%%%%%%%%%%%%%%%%%%%%%%%%%%%%%%%%%%%%%%%%%%%%%%%%%%%%%%%%%%
\caption{\label{fig6}{\small Chiral extrapolation of $\widehat g_Q^{\,(0)}$ for a
fixed heavy quark $\kappa_Q=0.1190$ at $\beta=6.0$ and $\kappa_Q=0.1250$ 
at $\beta=6.2$ corresponding approximately to the same heavy meson mass.}}
\end{center}\end{figure}
%%%%%%%%%%%%%%%%%%%%%%%%%%%%%%%%%%%%%%%%%%%%%%%%%%%%%%%%%%%%%%%%%%
%%%%%%%%%%%%%%%%%%%%%%%%%%%%%%%%%%%%%%%%%%%%%%%%%%%%%%%%%%%%%%%%%%
\begin{table}[t!!] 
\vspace*{.8cm}
\begin{center} 
\begin{tabular}{|c|c|c|c|c|c|}
\hline
{\phantom{\Huge{l}}}\raisebox{-.2cm}{\phantom{\Huge{j}}}
\hspace*{-5mm}&0.1339& 0.1342&0.1344&0.1346&$\kappa_{ud}$ \\
\hline
{\phantom{\Huge{l}}}\raisebox{-.2cm}{\phantom{\Huge{j}}}
\hspace*{-5mm}$G_1(0)$&$19.9\pm 2.4$&$20.8\pm 3.0$&$22.7\pm 4.3$&
$19.3 \pm 4.0$ & $21.4 \pm 6.0$ \\
{\phantom{\Huge{l}}}\raisebox{-.2cm}{\phantom{\Huge{j}}}
\hspace*{-5mm}$\widehat g_Q^{\,(0)}$&$0.80(7)$&0.80(9)&0.83(12)& 0.78(13) &0.80(19) \\
\hline
\end{tabular} 
%\vspace*{.8cm}
\caption{\label{tab7}
\small{\sl Values of $G_1(0)$ and $\widehat g_Q^{\,(0)}$ at $\beta=6.0$ for one fixed
heavy quark: $\kappa_Q=0.1190$ and several light quarks.
The extrapolation to the physical pion mass is presented in the last
column.  
}}
\end{center}
\end{table}
%%%%%%%%%%%%%%%%%%%%%%%%%%%%%%%%%%%%%%%%%%%%%%%%%%%%%%%%%%%%%%%%%%

It is worth noticing that the results at $\beta=6.0$ agree within errors  
with those at  $\beta=6.2$. This is illustrated in fig.~\ref{fig6}
where we compare the case $\beta=6.0,~\kappa_Q=0.1190$ to $\beta=6.2,~\kappa_Q=0.1250$, 
which corresponds to approximately the 
same physical mass of the heavy-light meson ($m_P=1.77(11)~\gev$
and $1.75(9)~\gev$ respectively). Notice that the numerical results at $\beta=6.0$ are more noisy 
than those at $\beta=6.2$.

For this reason and also because the lattice spacing at  $\beta=6.0$ implies 
strong limitations on heavy quark masses, we did not attempt an extrapolation to the charm
and even less to the infinite mass limit.
Notice nevertheless that the charm mass region is almost reached with 
$\kappa_Q=0.1190$ in our setup.

%%%%%%%%%%%%%%%%%%%%%%%%%%%%%%%%%%%%%%%%%%%%%%%%%%%%%%%%%%%
\section{Physical results and discussion of errors}
%%%%%%%%%%%%%%%%%%%%%%%%%%%%%%%%%%%%%%%%%%%%%%%%%%%%%%%%%%%

\subsection{Systematic uncertainties}

\begin{itemize}
\item {\it Discretisation errors-I}: In our study we implemented the full 
${\cal O}(a)$ improvement of the Wilson QCD action and the axial current. 
As we discussed in the text, the improvement of the bare axial current 
does not influence the value of our $g_{D^\ast D\pi}$. As for the 
renormalization constant, we used the non-perturbatively determined 
value, including the coefficient $\widetilde b_A$, which ensures the 
elimination of the artifacts of ${\cal O}(a\rho)$.

\item {\it Discretisation errors-II}: Our calculation has been made 
at $\beta = 6.2$. With intention to study the ${\cal O}(a)$ effects, 
we have also performed the simulation at $\beta =6.0$ which has been 
summarized in the preceding section. The good agreement illustrated 
in fig.~\ref{fig6} shows a small discretization error.
One might wonder if the small decrease of $\widehat g$ from $\beta=6.0$
to $\beta=6.2$ is the sign of a systematic finite $a$ effect. This would then
point toward a continuum limit lower than the values quoted here. 
This difference might also simply be a statistical one since 
it is smaller than one standard deviation. 
 With only two values
of the lattice spacing it is not possible to try a systematic study
of the continuum limit and thus to discriminate between these hypotheses.
 Further studies at different values of $\beta$
are badly needed. From the experience about similar quantities one
might hope that the continuum limit will not be too different
from the result at $\beta=6.2$.

\item {\it Discretisation errors-III}: When interpolating to the charm quark,
{\it i.e.} to the $\bar D$-meson~(\ref{fitts}), we used the value of the lattice
spacing as obtained from the pion decay constant. If the lattice spacing is fixed
by the $\rho$-meson mass, the value of the $g_{D^\ast D\pi}$ remains practically
unchanged. This is not surprising since the slope of $g_{VP\pi}/m_H$ in $1/m_H$ 
is very small so that the slight change of the position of the $1/ \bar m_D$ 
does not make any visible impact on the final result (it increases by  $\approx
1\%$).

\item {\it Chiral extrapolations}: This source of uncertainty actually  dominates
our systematic error bars. This shows up also in the difference between linear,
quadratic and logarithmic fits. Indeed, at $\beta=6.0$ the  additional lightest
quark ($\kappa=0.1346$) has been added in order to reduce this error.  We decide
to take the  result of the linear extrapolation as our central value since our
data follow rather clearly that form. The difference between that central value 
(of the linear extrapolations) and the ones obtained through the quadratic and
the ``log'' fits will be included in the systematic uncertainty.

%%%%%%%%%%%%%%%%%%%%%%%%%%%%%%%%%%%%%%%%%%%%%%%%%%%%%%%%%%%
\begin{figure}[ht]
\vspace*{-.cm}
\begin{center}
\begin{tabular}{@{\hspace{-0.7cm}}c}
\epsfxsize10.0cm\epsffile{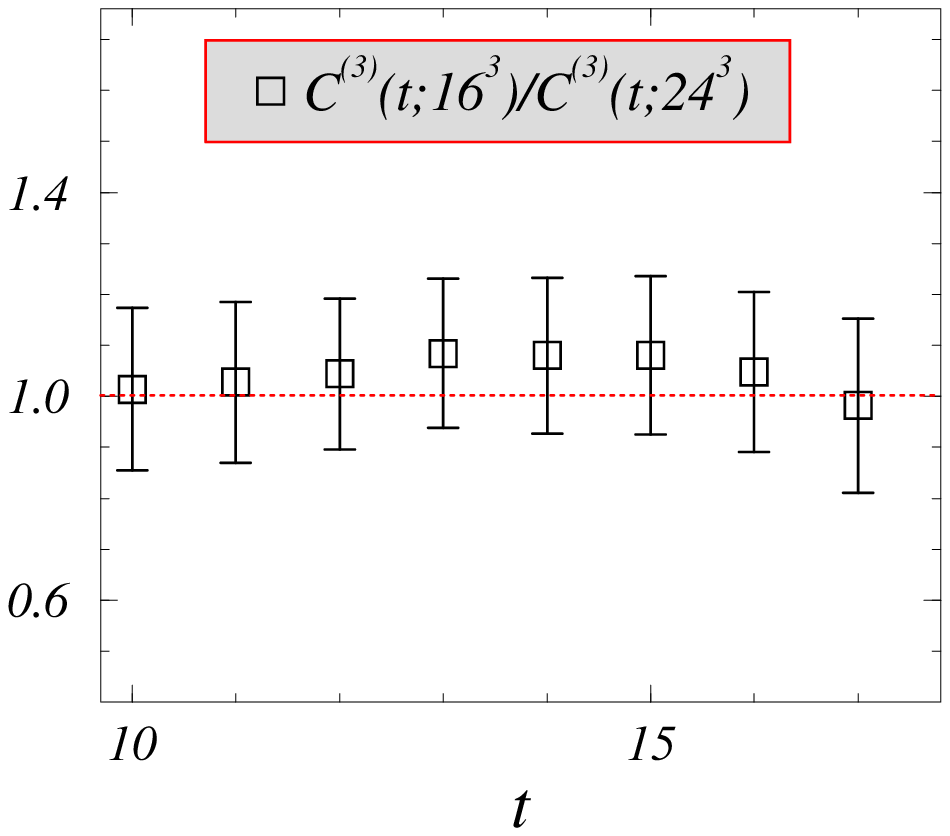}   \\
\end{tabular}
%%%%%%%%%%%%%%%%%%%%%%%%%%%%%%%%%%%%%%%%%%%%%%%%%%%%%%%%%%%%%%%%%%
\caption{\label{fig7}{\small Ratio of the three point functions~(\ref{eq1})
as obtained for the same meson masses at $\beta = 6.0$ but on different volumes 
$16^3\times 64$ and $24^3\times 64$. Illustrated is the case of the 
heavy quark corresponding to $\kappa_Q = 0.1220$ and the light ones to 
$\kappa_q=0.1344$.}}
%%%%%%%%%%%%%%%%%%%%%%%%%%%%%%%%%%%%%%%%%%%%%%%%%%%%%%%%%%%%%%%%%%
\end{center}
\end{figure}
%%%%%%%%%%%%%%%%%%%%%%%%%%%%%%%%%%%%%%%%%%%%%%%%%%%%%%%%%%%

\item {\it Finite volume effects}: We also studied the finite volume effect 
by performing two parallel simulations at $\beta =6.0$, with 
lattice of  size  $16^3\times 64$ and  
$24^3\times 64$. To illustrate the net effect, we plot in
fig.~\ref{fig7} the ratio of a three-point correlation functions~(\ref{eq1})
as obtained from the simulations with two lattice volumes and for both
heavy-light mesons being at rest. Within our statistics, we 
do not see any evidence for the presence of finite lattice volume 
effects. To be conservative, however, we will take into account the 
observation that  the central values are in the interval 
\bea
0.95 \leq {C^{(3)}_{ii}(t;16^3)\over C^{(3)}_{ii}(t;24^3)} \leq 1.07\;,
\eea
and thus will include $6~\%$ in the systematic error.

\item {\it Quenching effects}: 
When discussing the chiral extrapolations we also considered the 
 effect of using the leading chiral log behavior. It is important to
stress that such a behavior is only valid for the full (unquenched) QCD. 
In the quenched approximation, however, one encounters the so-called quenched
logs which are not of the form $m_p^2\log(m_p^2)$ as in full QCD but rather
divergent, of the form $m_0^2\log(m_p^2)$. Here $m_0$ stands for the mass of the
$\eta^\prime$-meson, which in the quenched theory does not decouple from the
octet of light pseudoscalar mesons. One could thus envisage a fit
to the form~\cite{sasa} 
\bea
g_{VPp} = c_0 + c_1 (a m_p)^2\log(a m_p) + c_1^{(quench.)} \log(a m_p)  
+ c_2 (a m_p)^2 + \dots 
\eea

If we had been in the region of  very small masses 
we should   have used this form of the fit to determine the coefficient 
$c_1^{(quench.)}$ and then correct for it when getting the quenched
physical results. 
In our case, however, the meson masses  we were able to
simulate directly are in the region in which the dependence of 
$g_{VPp}$ on $(am_p)^2$ is linear as can be seen in fig. 4.

\end{itemize}

\subsection{Conclusion}

In this paper, we made the first lattice QCD study of the strong  coupling of the
pion to the spin-doublet of $D$-mesons, $g_{D^\ast D\pi}$.  Our results, obtained
in  the quenched approximation, are in a very  good agreement with the recent
experimental measurement~\cite{CLEO}.  Our numbers are much larger than the
predictions made by  using various QCD sum rule techniques. It is also larger
than the  previous lattice estimate which has been made in the static limit of  
the heavy quark effective theory~\cite{UKQCD} and on a very coarse lattice.  The
reason for that disagreement remains to be understood.  It should be noted that
our data suggest that the dependence  on the heavy quark mass is very weak and
that the static value  $\widehat g_\infty \gtrsim \widehat g_c$.  A careful
extrapolation to the continuum  limit is still needed  but, inspired from
the experience with similar quantities, one might hope that
the continuum limit will not be too different from the result at $\beta=6.2$.

On the basis of our results and from the discussion of the systematic
uncertainties which we combine in the quadratic sum, we conclude that
\bea 
g_{D^\ast D\pi} = 18.8 \pm 2.3^{+1.1}_{-2.0} \,,\quad {\rm and}\quad \widehat g_{c} =
0.67 \pm 0.08^{+0.4}_{-0.6} \,,
\eea
which is the result that we quoted at the beginning of this
paper~(\ref{result62},\ref{eq:1}).

The coupling $\widehat g$ varies very little with the heavy mass and we find 
in the infinite mass limit 
\bea
\widehat g_\infty = 0.69 \pm 0.18.
\eea 

A further improvement of our results includes the increase of the 
statistical sample, computations with static heavy quarks, 
computations at larger $\beta$
(such as $\beta =6.4$), and most importantly the attempt of an unquenched 
study of this coupling.

\section*{Acknowledgement}

We thank  V.~Lubicz,  G.~Martinelli, S.~Prelovsek , J.-C. Raynal and J.~Zupan  
for precious discussions and comments. 
This work has been supported in part by the European Network 
``Hadron Phenomenology from Lattice QCD'', HPRN-CT-2000-00145.
 We have used for this work 
the APE1000 located  in the Centre de Ressources
Informatiques (Paris-Sud, Orsay) and purchased thanks to a funding from the
Minist\`ere de l'Education Nationale and the CNRS.

\end{document}